\documentclass[ prb, twocolumn, showpacs, superscriptaddress, amsmath,amssymb, letterpaper, balancelastpage, citeautoscript, 10pt]{revtex4}
\usepackage{amsmath,bm,hvfloat}
\usepackage[outercaption]{sidecap}
\usepackage{graphicx, dcolumn,graphicx,color,booktabs}
\usepackage{microtype}
\definecolor{DarkBlue}{rgb}{0, 0.0, 0.6}

\usepackage[colorlinks,plainpages=false,linkcolor=blue,urlcolor=blue,citecolor=blue,pdfpagemode=UseNone,pdfstartview=FitH]{hyperref}

\usepackage[charter]{mathdesign}
\usepackage[T1]{fontenc}
\definecolor{MyDarkBlue}{rgb}{0, 0.0,0.7}
\definecolor{MyRed}{rgb}{0.7, 0.0,0.0}

\begin{document}
\title{Pseudogap in the chain states of YBCO}

\author{V.\,B.\,Zabolotnyy}
\affiliation{Institute for Solid State Research, IFW-Dresden, P.\,O.\,Box 270116, D-01171 Dresden, Germany}
\author{A.\,A.\,Kordyuk}
\affiliation{Institute for Solid State Research, IFW-Dresden, P.\,O.\,Box 270116, D-01171 Dresden, Germany}
\author{D.\,Evtushinsky}
\affiliation{Institute for Solid State Research, IFW-Dresden, P.\,O.\,Box 270116, D-01171 Dresden, Germany}
\author{V.\,N.\,Strocov}
\author{ L.\,Patthey}
\author{T.\,Schmitt}
\affiliation{Paul Scherrer Institut, Swiss Light Source, CH-5232 Villigen PSI, Switzerland}
\author{D. Haug}
\author{C.\,T.\,Lin}
\affiliation{Max-Planck-Institut für Festkörperforschung, Heisenbergstraße 1, 70569 Stuttgart, Germany}
\author{V.\,Hinkov}
\affiliation{Quantum Matter Institute, University of British Columbia, Vancouver, B.C. V6T 1Z1, Canada}
\affiliation{Max-Planck-Institut für Festkörperforschung, Heisenbergstraße 1, 70569 Stuttgart, Germany}
\author{B.\,Keimer}
\affiliation{Max-Planck-Institut für Festkörperforschung, Heisenbergstraße 1, 70569 Stuttgart, Germany}

\author{B.\,B\"{u}chner}
\affiliation{Institute for Solid State Research, IFW-Dresden, P.\,O.\,Box 270116, D-01171 Dresden, Germany}
\author{S.\,V.\,Borisenko}
\affiliation{Institute for Solid State Research, IFW-Dresden, P.\,O.\,Box 270116, D-01171 Dresden, Germany}

\begin{abstract}

As established by scanning tunneling microscopy (STM) cleaved surfaces of the high temperature superconductor
YBa$_2$Cu$_2$O$_{7-\delta}$ develop charge density wave (CDW) modulations in the one-dimensional (1D) CuO chains.
At the same time, no signatures of the CDW have been reported in the spectral function of the chain band previously studied
by photoemission. We use soft X-ray angle resolved photoemission (SX-ARPES) to detect a chain-derived surface band that had not been detected in previous work.  The $2k_\textup{F}$ for the new surface band is found  to be 0.55\,\AA$^{-1}$, which matches the wave vector of the CDW observed in direct space by STM. This reveals the  relevance of the Fermi surface nesting  for the formation of  CDWs in the CuO chains in YBa$_2$Cu$_2$O$_{7-\delta}$.  In agreement with the short range nature of the CDW order the newly detected surface band exhibits a pseudogap, whose energy scale also corresponds to that observed by  STM.
\end{abstract}

\preprint{\textit{xxx}}

\maketitle
The CuO chain structure in YBa$_2$Cu$_2$O$_{7-\delta}$ (Y-123) provides a physical realization of a quasi-one-dimensional electronic system with non-vanishing coupling to the CuO$_2$ bilayers\cite{Andersen19951573, Pickett8764}. One of the reasons why 1D electronic systems remain in the focus of solid state research is that even weak interactions transform the quasi-particles of Fermi liquid theory into collective excitations of density wave type\cite{Tomonaga1950, Luttinger1963, Meden15753}. In particular, the peculiar  topological structure of the chain Fermi surface (FS) makes this electronic system prone to  formation of charge density wave via the Peierls instability\cite{Gruener}. Indeed,  charge density modulations along the Cu-O chains have been extensively studied probing  cleaved surfaces of Y-123 in direct space with scanning tunneling microscopy\cite{Edwards1154, Edwards1387,Edwards2967, Maki1877, Maki140511, Maki024536, Maki200284, Derro097002, Urbanik483}.
It has been shown that Y-123 crystals cleave between the CuO chains and the BaO layer, so that the chains turn out to be the nearest to the surface building block\cite{Edwards2967, Edwards1154, Maki1877, Urbanik483}. Both  the earliest\cite{Edwards2967} and the most recent studies\cite{Derro097002} present a consistent picture of a CDW, appearing as corrugations of the electronic density with a short correlation range of about 40\,{\AA} and a period between 9 and 14\,{\AA}, depending on the sample stoichiometry.

In the case of 2D systems, the effect of CDW on the electronic states in the reciprocal space has been examined in great detail with modern angle resolved spectroscopy, which provides both momentum and energy resolution when measuring the one particle spectral function. Occurrences of the pseudogap for an incommensurate (or short range ordered) state, which finally develops into a true CDW band gap below $T_\textup{CDW}$, are well documented\cite{Bovet125117, Yokoya140504, Borisenko196402,Borisenko166402}.  In the case of 1D systems other than Y-123, modifications to the spectral function with the onset of the CDW state have been detected as well \cite{Schafer066401, Koitzsch113110, Altmann035406}.
the emergence of a pseudogap-like state is also expected in theory and can be understood as a result of a fluctuating Peierls order parameter \cite{Bartosch988, Bartosch15488, Bartosch16223, Millis12496, Bartosch799}.

At the same time, despite the abundant evidence for charge modulations in direct space provided by STM,  Y-123 chains seem to exhibit neither a pseudogap nor the Tomonaga--Luttinger behavior observed in the related
 PrBa$_2$Cu$_4$O$_8$\cite{Mizokawa12335, Mizokawa193101}.  Momentum resolved spectra measured from as-cleaved Y-123
surfaces\cite{Nakayama014513, Zabolotnyy064519, Zabolotnyy024502}, as well as those measured on the \emph{in situ} doped
ones\cite{Hossain527}  manifest neither folding nor notable suppression of the spectral weight at the Fermi level. That is, the spectral features that could be regarded as signatures  of the CDW state are absent in spectra of Y-123. Therefore, it remains unclear why the two complementary methods
(STM and ARPES) deliver such conflicting results.

To address this problem we have investigated high quality Y-123 crystals using modern SX-ARPES.
the photoemission data for this study were collected at the recently built high-resolution soft X-ray beamline ADRESS at the Swiss Light Source \cite{Strocov631}. This beamline delivers photon flux  up to 10$^{13}$\,ph/s in an energy window of 0.01\% of photon energy. Such a high flux allows one to  break through the problem of dramatic reduction of the valence band cross section at high photon energies. the samples were mounted on a low-temperature goniometric manipulator (CARVING) with three angular degrees of freedom and cleaved in situ in ultrahigh vacuum with base pressure better than 5$\cdot$10$^{-11}$\,mBar at $T$=10\,K, the same temperature at which all the spectra were acquired. The energy resolution depends on the kinetic energy and will be stated in the figure captions, details of the SX-ARPES station will be published elsewhere\cite{Strocov}.
The experiments were performed using high-quality single crystals of Ba$_2$Cu$_3$O$_{6.6}$ ($T_\textup{c}\!\thickapprox$61\,K)  from the same batch as in the recent INS study\cite{Hinkov650}. The crystals  were synthesized by the solution-growth technique, annealed to the desired oxygen doping, and detwinned by applying uniaxial mechanical stress at elevated temperature\cite{Hinkov650}.

As a result of the `surface polar catastrophe'\cite{Hossain527} the electronic structure of the Y-123 surface layer is known to be different from that of the bulk\cite{Pasanai134501}.    Generally, there are two major approaches that have been used in photoemission practice to highlight   bulk states against the surface ones. Surface aging is the first option, which basically relies on the destruction of surface sates\cite{Lu4370, Damascelli5194}. The other approach employs the variation of photon excitation energy and polarization, which affect the inelastic escape depth of exited electrons\cite{Seach1979, Powell19991, Tilinin547} and the ratio of photoemission matrix elements for bulk and surface states\cite{Hansen1871, Zabolotnyy024502}.
Also the rapid variation in the elastic scattering rate with the kinetic energy of the excited photoelectrons may play a decisive role\cite{Krasovskii165406, Barrett035427}.

\begin{figure}[t]
\includegraphics[width=0.9\columnwidth]{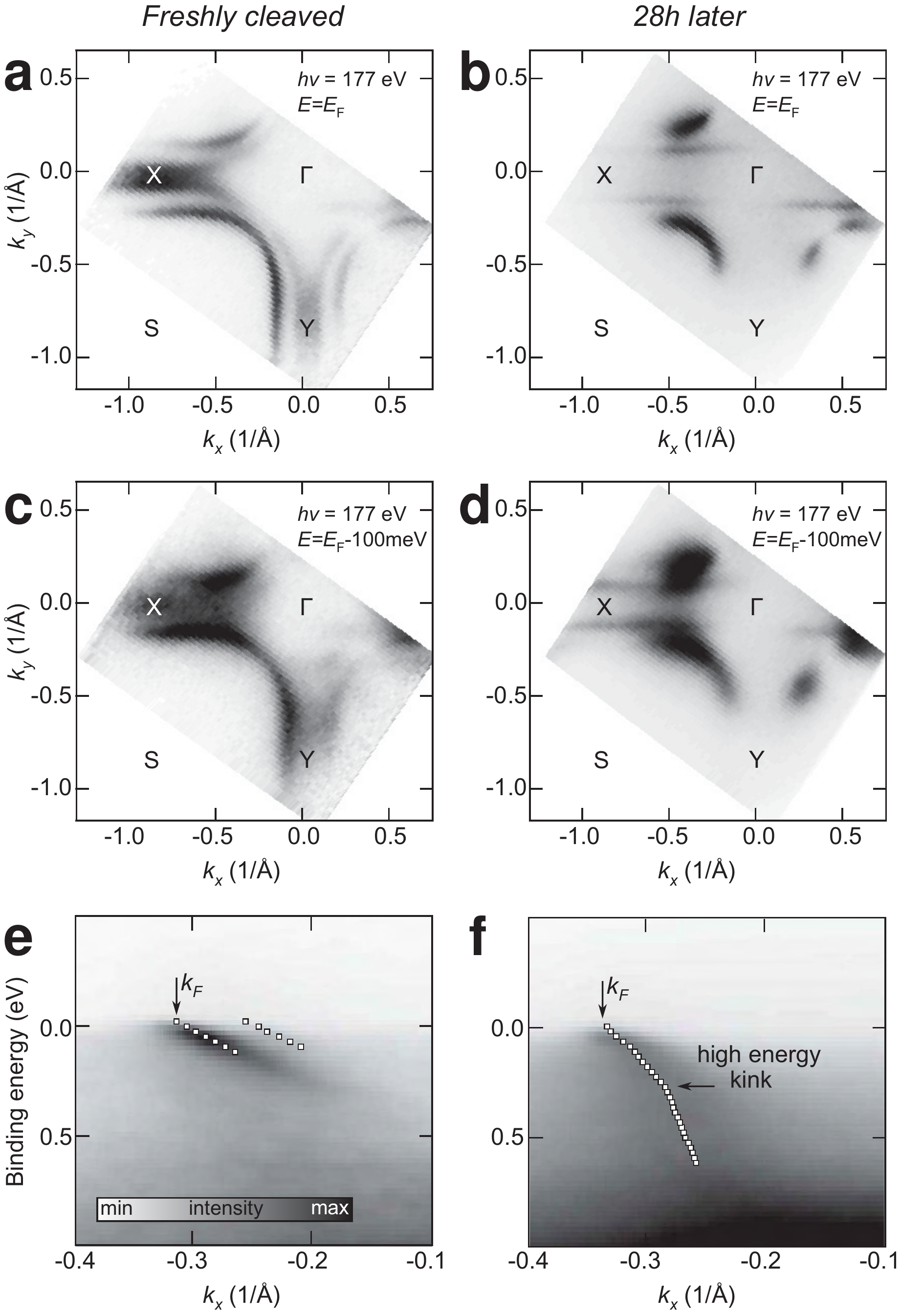}\\
\caption{Sample aging. (a), (b)  FS maps measured directly after cleavage and 28 hours later.
 (c), (d) Corresponding fresh and aged intensity distributions for constant energy cuts at $E=E_\textup{\tiny{F}}-100$\,meV.
(e), (f) Energy--momentum intensity distributions along the $\Gamma$--\,S direction. The white squares are the result of MDC fits.  Dark corresponds to high photoemission intensity. Spectra measured at $T=10$\,K with 38\,meV energy resolution using linearly $s$-polarised light (i.e. no out of plane component of the polarization vector).  }
\label{aging}
\end{figure}

We start out the discussion of experimental data  with an illustration of the first option. Fig.\,\ref{aging} contains a comparison of the Fermi surfaces measured from the freshly cleaved sample to the same measurement done 28 hours after the cleavage. The freshly cleaved surface results in a typical picture of bilayer split FS contours corresponding to the CuO$_2$ plane states (rounded double squares centered at S point) and the quasi-one-dimensional chain band (faint features running parallel to the $k_x$ axis).  After surface degradation there are some notable changes.

First,  the overdoped CuO$_2$ plane bands are replaced with the `Fermi arc' features similar to those observed in  refs.\,\onlinecite{Fournier905, Sassa140511}. The appearance of the `Fermi arcs' is also accompanied
by a $k_\textup{\tiny{F}}$ shift (Fig.\,\ref{aging}(e--f)).   For the bonding band, it can be estimated from the MDC fits, and amounts to about $\delta \!k_\textup{\tiny{F}}\approx$ 0.04\,{\AA} along the $\Gamma$--\,S direction. Assuming the same shrinking over the whole Fermi surface and an average FS radius $k_\textup{\tiny{F}}\approx0.55$\,{\AA}, one may estimate the decrease in hole doping as
$ \delta \!p \approx 4\pi k_\textup{\tiny{F}} \delta \!k_\textup{\tiny{F}} / S_\textup{\tiny{BZ}} \approx 0.1$,  which brings us from the typically overdoped  regime with $p\approx0.3$ (refs.\,\onlinecite{Nakayama014513, Zabolotnyy2007888, Zabolotnyy064519, Lu4370}) under the superconducting dome with $p\approx0.2$. Note also the so called `waterfalls' and a high energy kink at about 250\,meV, which have recently generated an avalanche of publications \cite{Katagiri165124, Basak214520, Inosov237002, Inosov212504}.  These  issues are beyond the scope of the current study, and we do not further elaborate on them here.

Second, the relative intensity of the 1D chain band as compared to the 2D bands is substantially increased, while the distance between the chain Fermi crossings, $2k_\textup{F}$, remains practically the same.
This suggests that the observed chain band is most likely a `bulk' feature, in a sense that
it originates from the chain structure protected by at least one CuO$_2$ bilayer, and not from the cleaved chains.
Since the upper neighboring CuO$_2$ bilayers  turn out to be  overdoped, the electronic structure of these chains still differs from  LDA band structure calculation\cite{Pickett8764}, so  in the following we will refer to these chains as subsurface ones,
in order to contrast this structure with the chains through which the cleavage takes place and  the true bulk chains.
\begin{figure}[b]
\includegraphics[width=0.9\columnwidth]{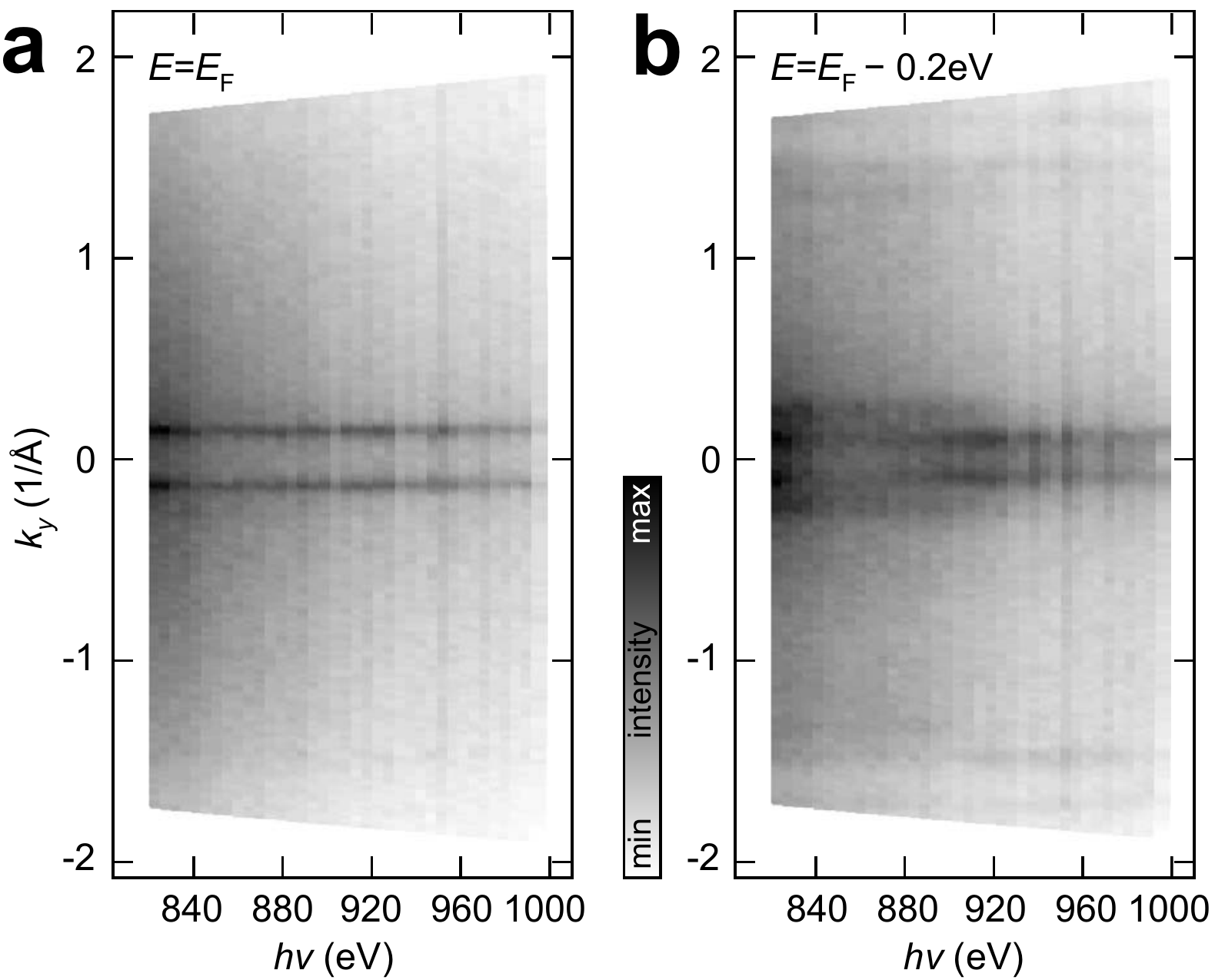}\\
\caption{Excitation energy dependence for the cut passing through the $\Gamma$ point parallel to the $k_y$ axis. The left panel shows the intensity integrated around the Fermi level, $E=E_F$. The right image contains intensity  at 200\,meV below the Fermi level, demonstrating another set of chain features for $h\nu \lesssim$920\,eV.  Spectra were measured using $p$-polarized light with overall  energy resolution of 125\,meV.}
\label{e-dependence}
\end{figure}

This observation is consistent with the interpretation given in ref.\,\onlinecite{Zabolotnyy024502}, which was based on the pattern of circular dichroism.  This work has shown  that the chain signal observed in ARPES experiment mainly originates from the nearest to the surface  \emph{undisturbed} chains protected by the overdoped CuO$_2$ bilayer, so that  the surface aging is expected to enhance the photoemission from these chain states as compared to other surface features.
 Though there are intensity variations along the chain band, it is noteworthy that the band
 exhibits  neither  pseudogap  nor  folding phenomena.  As can be seen in Fig.\,\ref{aging}(a--d), there are no indications for a CDW state for the freshly cleaved and aged surfaces.

Now we turn to the discussion of the second option.
Searching  for the signatures of a  CDW in Y-123 spectra we have measured the excitation energy dependence for the Y--$\Gamma$--Y cut, the one where only a  parabolic chain band is expected to
 cross the FL\cite{Pickett8764}. Fig.\,\ref{e-dependence} contains the intensity distribution for that direction plotted at the FL (a)
 and 200\,meV below the FL (b). While the data in panel (a) exhibit only two parallel features, corresponding to the $\pm k_\textup{F}$
 crossings of the two branches  of the parabolic chain band, the intensity distribution in Fig.\,\ref{e-dependence}(b) reveals another set of features  running parallel  with a slightly larger  separation. The new features are well visible  at low excitation energies $h\nu \lesssim$920\,eV, therefore based on the universal dependence of the escape depth one may assume the new feature to be a surface related one.
However, the decrease of the photoemission intensity from the surface chain band as compared to the  subsurface ones   occurs rather abruptly, whereas the universal escape depth curve predicts smooth $\propto E^{1/2}_\textup{kin}$ dependence, where $E_\textup{kin}$ is the kinetic energy of the excited photoelectrons. Therefore this observation cannot be solely attributed to the electron escape depth, so  also the $h\nu$ dependence of the photoemission matrix elements needs to be taken into account. In contrasting surface photoemission with bulk emission, particular attention has to be payed to the surface induced light fields, which are likely to be responsible for the different $h\nu$ dependence of the matrix elements for surface and bulk localized states.\cite{Levinson628, Meyers2030, Feuerbacher0022}  For instance, the absence of the new surface chain states in the spectra shown in Fig.\,\ref{aging} would be consistent with the fact that these spectra were measured with $s$-polarized light. For this polarization there is no component of the exciting field $\mathbf{A}(\mathbf{r})$ perpendicular to the sample surface, and hence the surface related term in the photoemission matrix element
$\langle \textup{f}| \textbf{A}\nabla+\frac{1}{2}\operatorname{div}\!\textbf{A}| \textup{i} \rangle$  is negligible, rendering
the surface emission effects irrelevant as compared to the case of $p$-polarized light.

\begin{figure}[t]
\includegraphics[width=1\columnwidth]{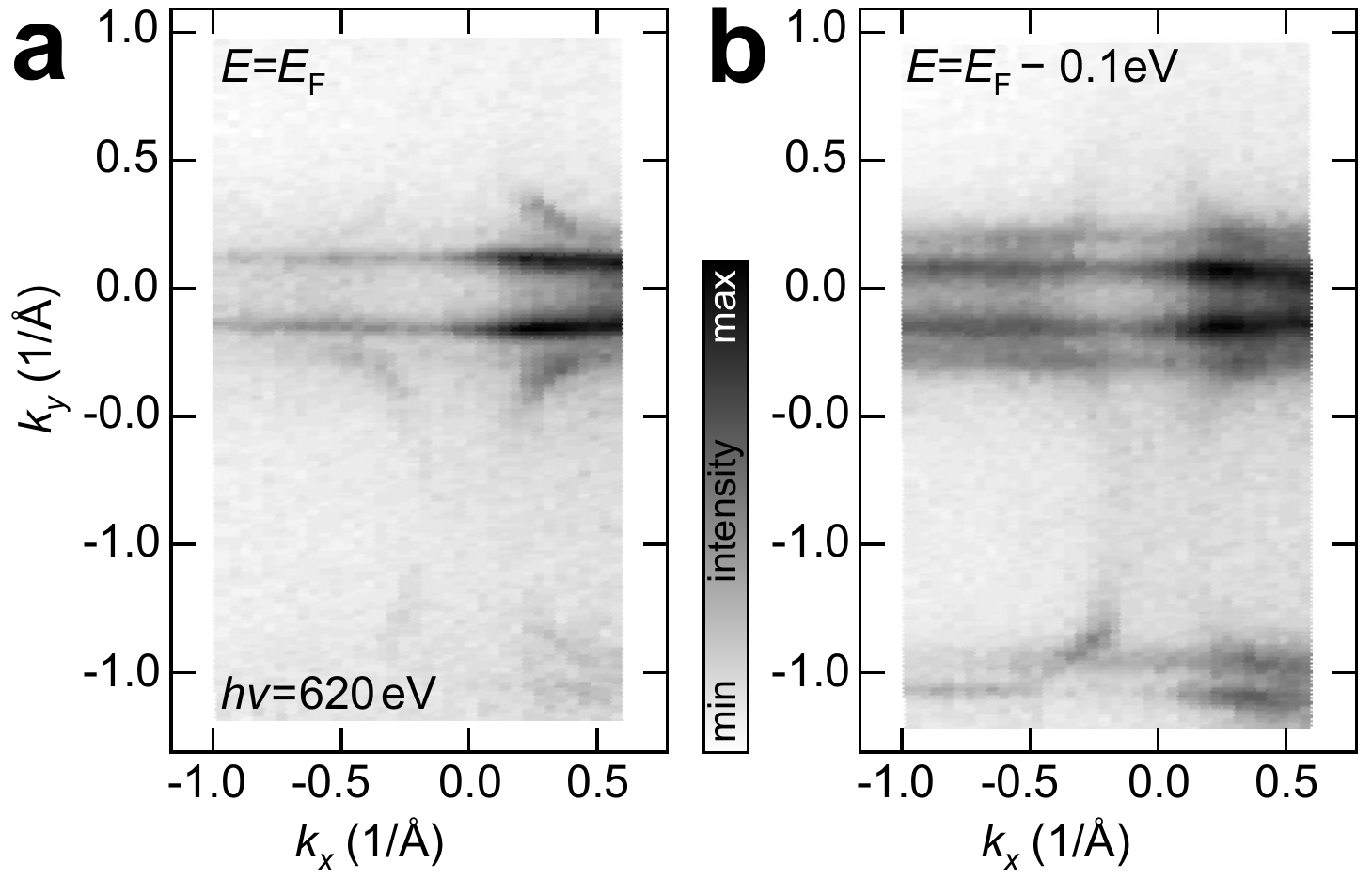}\\
\caption{(a), (b) Momentum intensity distribution at the Fermi level and at 0.1\,eV below the FL.  The right image
reveals another 1D band,  which contributes practically no spectral weight at the Fermi level.
Spectra measured with $p$-polarized light and energy resolution of 90\,meV.
}
\label{pseudogapMap}
\end{figure}

To establish the dimensionality and topology of this previously overlooked surface state,  in Fig.\,\ref{pseudogapMap}(a,b)  we plot a FS measured with 620\,eV photons and corresponding iso-energy intensity distribution taken 100\,meV below the FL.  According to the $hv$ dependence, the new feature is expected to have significant contribution to the experimental spectra at this energy. As can be seen, the new band practically does not contribute spectral weight at the FL (pseudogapped), though at 100\,meV binding energy the band gives rise to two 1D traces similar to those spawned by the chain band. This suggests that  the subsurface chains and the pseudogapped surface feature stem from homologous bands with, probably, different spatial localization. Indeed, a detailed analysis of the LDA band structure\cite{Pickett8764} shows that there are no other 1D bands except for the chains that the pseudogapped feature could be attributed to. Further, $k_z$ dispersion cannot be responsible for the appearance of the second pseudogapped chain band in the spectrum, since along the  Y--$\Gamma$--Y  direction
the calculated $k_{\textup{F}\parallel}$ variation with $k_z$ ($\pm$4\% $k_{\textup{F}\parallel}$) is notably smaller than the observed splitting.

In Fig.\,4(a, b) we plot the energy--momentum intensity distribution for the Y--$\Gamma$--Y direction. Panel (b) contains the same data set, but each MDC making up the image  has been normalized to a fixed value, in order to show the band dispersion in the  region that is `overexposed' in panel (a).

\begin{figure*}[t]
\label{pseudogap}
 \flushright
\begin{minipage}[l]{0.66\linewidth}
 \includegraphics[width=\textwidth]{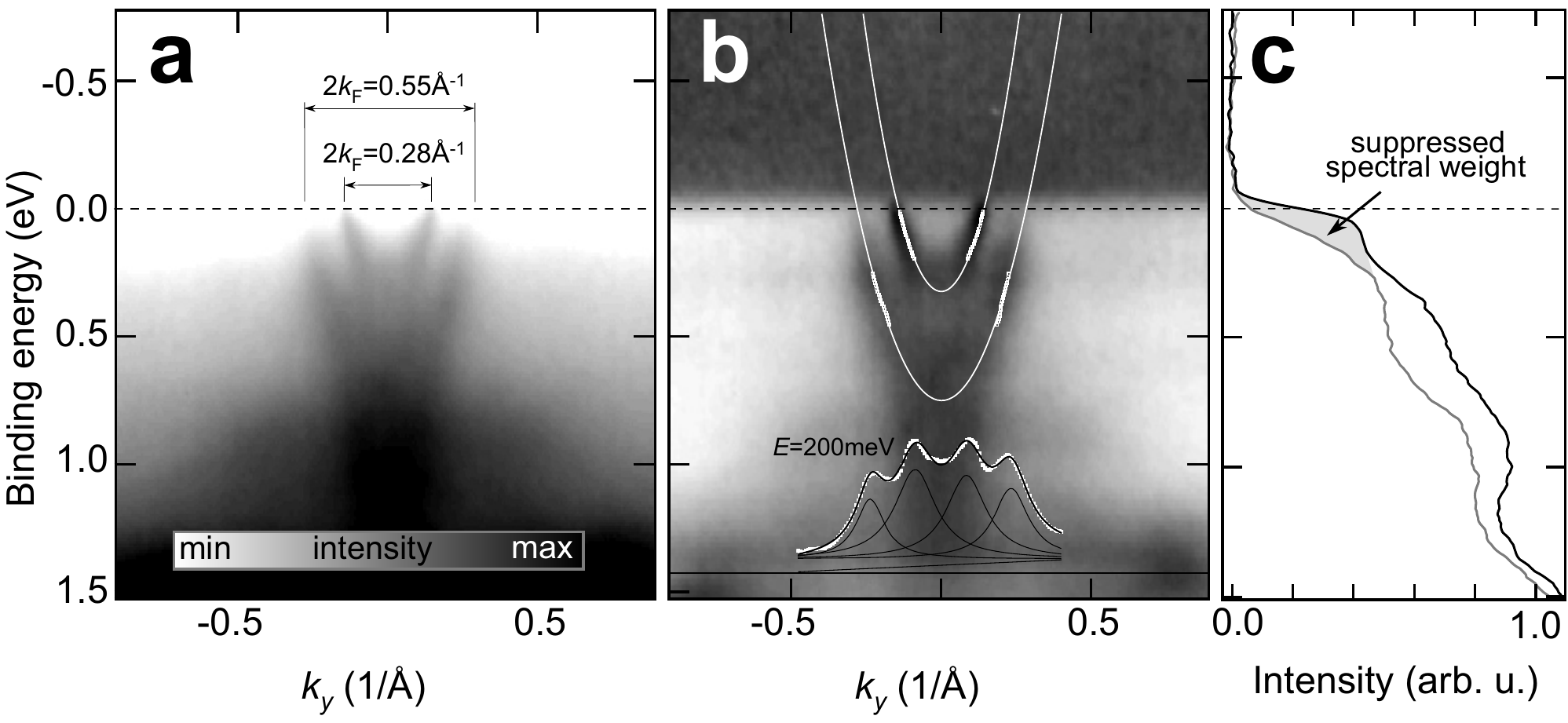}
  \label{pseudogap}
 \end{minipage}
 \hspace{0.015\linewidth}
 \begin{minipage}[r]{0.30\linewidth}
     \vspace{-0.6cm}
     \caption{(a) Energy-momentum cut along the Y--$\Gamma$--Y direction demonstrating two types of chain bands.
    (b) The same as previous, but each MDC in the image has been normalized to one.
    The small square symbols denote the peak positions obtained in a four-Lorentzian MDC fit. The insent
    shows typical fit for the MDC at $E_\textup{B}$=200\,meV.  The obtained experimental dispersions are fit to
    parabolic bands (solid white lines). (c) $k_\textup{F}$ EDC's for the inner chain band (black line) and outer chain band (grey line). Polarization and resolution are the same as in Fig.\,\ref{pseudogapMap}.}
    \label{pseudogap}
\end{minipage}
\label{pseudogap}
\end{figure*}

    The inner band with $2k_\textup{F}\sim$0.28\,{\AA}$^{-1}$ is the chain band that we attribute here
    to the subsurface chain states, and the outer one,  with $2k_\textup{F}\sim$0.55\,{\AA},$^{-1}$ is the newly
    observed band, which we believe  to be a surface localized chain band. The suppression of the spectral
    weight down to about 250\,meV for the surface chains as compared to the bulk ones is clearly visible.
    In the panel (c) we additionally compare two EDC's taken at the effective $k_\textup{F}$ of the bulk and
    surface bands, which once again demonstrate the pseudogap-like suppression of the spectral weight for
    the surface band.  While for the subsurface chains the spectral intensity grows approximately as the Fermi function
    convoluted  with experimental resolution,  for the surface band a gradual growth of intensity is observed. The
     estimated energy scale is about 200--300\,meV.

In the case of true CDW gap, using model calculations,  it has been demonstrated that new states arise on the order of $E_\textup{F}\pm\Delta$.  For these energies, the CDW component of the local DOS was shown to have a characteristic amplitude and phase.\cite{SacksS925}  Extending this results on the pseudogap, we note that the pseudogap nicely agrees with the STM results of ref.\,\onlinecite{Edwards1154}
and ref.\,\onlinecite{Maki024536}. Namely, the charge corrugations are visible for bias voltages up to 310\,meV and practically vanish for bias voltages higher than 480\,meV.   In this regard we can also conclude that another small gap ($\sim20$\,meV)
observed in STM spectra and attributed either to CDW gap or to proximity induced superconducting gap\cite{Edwards1387},
is likely  to be a superconductivity related one.

It is also informative to see how the observed  band dispersion compares to the CDW modulations measured in STM experiment.
Simple theory predicts that the CDW wave vector should be twice the Fermi vector,
$q_\textup{CDW} = 2\pi / \lambda_\textup{CDW} =  2k_\textup{F}$.   To estimate from the photoemission spectrum where
the pseudogapped chain band would cross the FL, if there were no CDW instability, we use the result of an MDC fit performed
for energies below the gapped region (white symbols in Fig.\,4(b) and extrapolate the experimental points
up to the Fermi level assuming a parabolic dispersion. The value thus obtained is $2k_\textup{F}=0.55$\,\AA$^{-1}$, which yields  $\lambda_\textup{CDW}  = 11.4$\,\AA. This value nicely compares to the $\lambda = 11.2$\,\AA, observed in
Zn substituted Y-123\cite{Maki024536}.  It is also within the range of CDW periods from 9 to 14\,{\AA} reported
for a series of YBa$_2$Cu$_2$O$_{7-\delta}$ samples with varying stoichiometry\cite{Maki140511}, approximately corresponding
to the sample with $\delta=0.35$ used in that study.

At the same time for the subsurface chains with $2k_\textup{F}=0.28$\,\AA$^{-1}$, so the expected CDW period would be about 22\,{\AA},
which obviously does not mach the STM data. Together with the absence of the (pseudo)gap this once again suggests that
CDW modulations, which have been studied in so many details by STM,  are due to the $2k_\textup{F}=0.55$\,\AA$^{-1}$ instability
occurring in the previously overlooked surface chain band.  Disorder in the chain structure results in a short range character of the CDW and the development of the pseudogap in the chain band.

It is remarkable that the chains at the surface develop the CDW state, while the neighboring chains sandwiched between the overdoped and bulk CuO$_2$ layers display no notable signatures of the CDW.  One reason for this could be a trivial difference in phonon modes available at the surface and in the bulk.  However commensuration effects appear to be of no less importance.  As compared to the incommensurate state, there is an additional energy lowering associated
with the commensurability of the CDW state, which is not taken into account in the simplest variant
of the CDW  theory\cite{Gruener, Lee1974703}.  The energy gain is given by $E_\textup{\tiny{comm} } = -\frac{n(\epsilon_\textup{\tiny{F}}) \Delta^2 }{\lambda_\textup{\tiny{el--ph}}}\left(\frac{\Delta}{D}\right)^{M-2}$. Here $D$ is the band width,  $\Delta$ is the CDW gap, $\lambda_\textup{\tiny{el--ph}}$ is the  electron--phonon coupling  constant, $n(\epsilon_\textup{\tiny{F}})$ - density of states at the Fermi level, and $M = \lambda\textup{\tiny{CDW}} / a$ is the commensurability factor. As can be seen the correction is most significant at low $M$  values. Considering this correction, for the surface chains we have $M \approx 11.4/3.89\approx 2.94$, which is indistinguishable from a commensurate modulation with $M=3$. For the subsurface chains $M=5.77$ is quite large and  differs significantly from the nearest commensurate value $M=6$. Consequently the $E_\textup{\tiny{comm} }$ correction is negligible in this case, which obviously should affect the energy balance, making the  CDW phase less favorable for the subsurface chains.

In conclusion, we have identified the origin of the CDW observed at the surface of Y-123 by STM by detecting
a previously overlooked  1D surface band in Y-123, which brings into agreement two complementary methods: ARPES and STM.   As a result of a Peierls  instability, the surface CuO chain band develops a short range CDW state, which results in the appearance of a pseudogap in the one particle excitation spectrum detected in our ARPES experiment. Both energy and momentum scales measured in STM and ARPES are found to be in agreement with each other.

We thank F. Dubi and C. Hess  for technical support during the measurements.
The project was supported in part by DFG grant ZA 654/1-1.
%

\end{document}